# BitWhisper: Covert Signaling Channel between Air-Gapped Computers using Thermal Manipulations

Mordechai Guri

Department of Information Systems Engineering
Ben-Gurion University
Beer-Sheva, Israel
gurim@post.bgu.ac.il

Matan Monitz, Yisroel Mirski, Yuval Elovici

Department of Information Systems Engineering
Telekom Innovation Laboratories at Ben-Gurion University
Beer-Sheva, Israel
{monitzm,yisroel}@post.bgu.ac.il, elovici@bgu.ac.il

*Abstract*— **It has been assumed that the physical separation ('air-gap') of computers provides a reliable level of security, such that should two adjacent computers become compromised, the covert exchange of data between them would be impossible.**

**In this paper, we demonstrate *BitWhisper*, a method of bridging the air-gap between adjacent compromised computers by using their heat emissions and built-in thermal sensors to create a covert communication channel. Our method is unique in two respects: it supports bidirectional communication, and it requires no additional dedicated peripheral hardware. We provide experimental results based on implementation of BitWhisper prototype, and examine the channel's properties and limitations. Our experiments included different layouts, with computers positioned at varying distances from one another, and several sensor types and CPU configurations (e.g., Virtual Machines). We also discuss signal modulation and communication protocols, showing how BitWhisper can be used for the exchange of data between two computers in a close proximity (at distance of 0-40cm) at an effective rate of 1-8 bits per hour, a rate which makes it possible to infiltrate brief commands and exfiltrate small amount of data (e.g., passwords) over the covert channel.**

 *Keywords—air-gap; bridging; covert channel; temperature; sensors; exfiltration; infiltration (key words)*

## I. INTRODUCTION

An air-gapped network is a computer network in which security measures are taken to maintain physical and logical separation from other, less secured networks. The term *air-gap* may sometimes refer to an interface between two systems or networks in which data transfers are handled manually (e.g., a human operator copies data from one system onto a flash-drive and then walks over to another system and connects the flash-drive to it [1]). Air-gapped networks are often used in cases in which the information stored or generated by the network is highly sensitive or at risk of data leakage. For instance, military networks such as the Joint Worldwide Intelligence Communications System (JWICS) are air-gapped networks [2] [3]. Despite the added security benefits of an air-gapped network, such networks have been breached in recent years. The most famous cases are Stuxnet [4] and agents.btz [5], although other cases have also been reported [6].

Despite these attacks, air-gapped networks are still used because they minimize the risk of data leakage and prevent malicious code and commands from being transmitted to the network.

### A. Related Works

Covert channels in general are widely discussed in professional literature, using diverse methods [7] [8] [9]. However, in this paper, we focus on covert channels that can bridge air-gapped – or physically separated – computers. This subset of covert channels employs several physical phenomena, utilizing acoustic inaudible channels, optical channels, and electromagnetic emissions. Madhavapeddy et al [10] discuss so-called 'audio networking', while Hanspach and Goetz [11] disclose a method for near-ultrasonic covert networking. Loughry and Umphress [12] discuss information leakage from optical emanations. Software-based hidden data transmission, using electromagnetic emission, is discussed by Kuhn and Anderson [13]. Guri et al [14] present 'AirHopper', a malware that utilizes FM emissions to exfiltrate data from an air-gapped computer. Note that the aforementioned channels are unidirectional and do not permit the establishment of an inbound channel into an isolated network. Technically speaking, thermal radiation is a form of electromagnetic emanation, and this was, in fact, previously suggested by Murdoch [15] for use as a covert channel (discussed in 'future work' section). To the best of our knowledge, our current paper is the first to present academic research with successful and detailed results in this respect.

### B. Our contribution

In this study, we present *BitWhisper*; a novel method for exchanging information between two adjacent personal computers (PCs) that are part of separate, physically unconnected networks. BitWhisper establishes a covert channel by emitting heat from one PC to the other in a controlled manner. By regulating the heating patterns, binary data is modulated into thermal signals. In turn, the adjacent PC uses its built-in thermal sensors to measure the environmental changes. These changes are then sampled, processed, and demodulated into binary data.

Our experimental results demonstrate that BitWhisper is capable of a rate of eight signals per hour. Although this rate may seem slow compared with the previously mentioned methods of bridging air-gaps, BitWhisper offers two unique and useful characteristics: 1) the channel supports bidirectional (half-duplex) communication as both PCs can act as a transmitter (producing heat) or receiver (by monitoring the temperature), and 2) establishing the channel is possible using off-the-shelf adjacent desktop PCs and requires no special hardware or supporting components. These properties enable the attacker to exfiltrate information from inside an air-gapped network, as well as transmit commands to it. Furthermore, the attacker can use BitWhisper to directly control a malware's actions inside the network and receive feedback.

The remainder of this paper is structured as follows. Section II discusses the attack scenario, and section III provides technical background. Sections IV and V present the experimental setup and experimental results respectively. Section VI discusses the data modulation. Section 0 discusses countermeasures, and we conclude with section VII.

## II. ATTACK SCENARIO

BitWhisper, as a generic covert channel, can be utilized for various purposes. However, we explore its use as a method for bridging the air-gap between physically separated networks. In this section we discuss a possible attack model in which BitWhisper is used by an attacker to achieve this goal. The environmental settings are comprised of a facility or office space in which PCs are either part of an isolated network or a public network, such as the Internet. The attack model consists of several phases. During the first phase, the attacker infects networks connected to the Internet, a part of the attack which can be accomplished using targeted malicious emails combined with social engineering and other similar methods. During the second, considerably more challenging phase, the attacker goes on to contaminate a node of the internal network. This can be done by attacking the supply chain [16], planting an infected USB drive, using a malicious insider, or some equivalent tactic. Several recent incidents have shown that this type of breach is feasible [4] [5] [6] [17].

At this point, having established a foothold in both networks, the attacker may bridge the air-gap between the networks in order to covertly exfiltrate a highly sensitive piece of information (e.g., passwords, or secret keys). Alternatively, the attacker may trigger a worm attack inside the isolated network or send a malicious command to an isolated industrial control system. After infecting the networks, the malware spreads over both networks and searches the surroundings for additional PCs within close proximity, spatially. Proximity is determined by periodically sending 'thermal pings' (demonstrated in Figure 1) over the air. Once a bridging attempt is successful, a logical link between the public network and the internal network in established. At this stage, the attacker can communicate with the formerly isolated network, issuing commands and receiving responses.

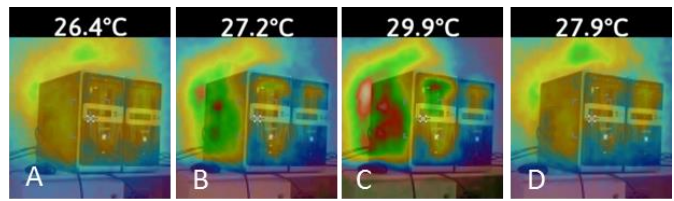

**Figure 1. A "thermal ping" sent between two adjacent PCs. The snapshots were taken by using a thermal camera.**

## III. TECHNICAL BACKGROUND

In this section, we provide a technical survey of the thermal emissions of PCs along with methods used to cool them. This review will assist the reader in understanding how heat from one PC is emitted and then detected by an adjacent PC in a controlled manner.

### A. Ambient Heat Sources in a PC

Like many electrical systems, PCs generate heat. The law of conservation of energy states that energy is conserved over time; excess power dissipates as heat, primarily in a physical process called Joule heating. Joule heating (also termed resistive heating) occurs when the passage of an electric current through a conductor releases heat. The generated heat is proportional to the current and voltage of the system [18]. Complex electronic systems such as the central processing unit (CPU) of a modern PC requires varying amounts of power (current and voltage) proportional to the workload of the system. This workload directly affects the amount of heat generated by the system. The dynamic power consumption of a CPU is generalized by the formula

$$P \cong \propto CfV^2 \quad (1)$$

Where $P$ is the power consumption in watts, $C$ is the capacitance of the CPU in farads, $f$ is the operating frequency in Hz, $V$ is the voltage in volts and $\propto$ is the percentage of the system that is active [19]. When the system is under load and executing different instructions, $\propto$ clearly increases as does the capacitance $C$, because more switching gate capacitors need to be charged. Higher operating frequencies also require quicker charging and discharging of logical gates, achieved by increasing the CPU voltage and frequency. To summarize, an increased CPU workload on a modern system causes increased power consumption across all parameters, therefore increasing the temperature as well.

In addition to the CPU, modern computers also contain other components that emit a significant amount of heat, including the graphics-processing unit (GPU) and other motherboard components such as the voltage regulator modules (VRM) and I/O controllers- Other heat sources in a computer include mechanical systems such as a hard drive or optical drive.

### B. Thermal Sensors in PCs

Electronic systems like computers and smartphones incorporate multiple thermal sensors to monitor their various components and ambient temperature. Thermal sensors allow the system to protect itself from damage or performance degrada-tion by decreasing the system's workload, activating external cooling systems such as fans, or even performing emergency shutdowns. In this section, we briefly survey common thermal sensors found in modern computer systems. Later, in section VI, we use some of these sensors to receive thermal signals transmitted through the air.

*1) CPU/GPU Thermal Sensors*

CPUs and GPUs contain multiple built-in temperature sensing diodes. In the case of CPUs, these diodes enable the system to measure the internal temperature of each core of the CPU separately. In addition to the CPU core's internal sensors, some processors include sensors that monitor the outside of the CPU package. In a typical PC, the difference between the internal temperature of a CPU and the ambient temperature inside the PC's chassis may be tens of degrees. Interestingly, CPU sensors are designed to be more accurate at higher operating temperatures, while being notoriously inaccurate at lower temperatures [20].

*2) Motherboard Thermal Sensors*

Motherboards usually incorporate thermal sensors and cooling fan controls in order to allow for buzzer alarms or emergency shutdowns at extreme temperatures, thus preventing damage to the computer's sensitive components. Sensing is usually performed by the I/O controller for easy monitoring by an application. It is common to place these sensors in locations that are relatively distant from heat sources, in order to monitor the motherboard's temperature itself. In many motherboard models, the overheat threshold (the CPU temperature at which the computer shuts down to prevent damage) can be changed in the BIOS.

*3) Other Thermal Sensors*

Voltage regulators in a PC supply the required voltage to its CPUs and GPUs. Voltage converters are generating a significant amount of heat and require thermal sensing and control. Some computer cases have fans and other cooling methods intended to cool a specific component while dissipating heat from the entire chassis. To that end, thermal sensors are placed in the way of the airflow created by these fans, or on the chassis itself. Most hard disk drives and solid-state drives have built-in temperature sensors. A diagnostic software can access these sensor readings through an interface called the Self-Monitoring, Analysis and Reporting Technology (S.M.A.R.T) interface [21].

Table 1 summarizes the thermal sensors and their primarily characteristics.

**Table 1. Survey of thermal sensors (common values)**

| sensor | idle temperature (°C) | max temperature (°C) | sensor resolution |
|---|---|---|---|
| CPU | 33-35 | 130 | 1°C degree |
| GPU | 33-35 | 105 | 1°C degree |
| Motherboard | 33-35 | 90 | 1°C degree |
| Voltage regulator | 31-33 | 90 | 1°C degree |
| Hard-disk drives | 35 | 55 | 1°C degree |
| Solid-state drives | 25-30 | 70 | 1°C degree |

*C. Cooling Mechanisms for PCs*

Some components of a computer may become temporarily unusable or damaged permanently if not cooled properly. For stable continuous operation of the PC, these components must receive sufficient cooling to counter their heat emissions. There are two methods of cooling: passive and active.

**Passive methods** cool the component by letting the heat dissipate naturally into the air on its own or with the help of a heat-sink (to conduct and disperse the heat via direct contact). For most chips found in a PC, this is the cooling method of choice. However, some chips may generate enough heat under regular usage to damage themselves. In these cases, passive cooling is not sufficient.

**Active methods** attempt to speed up the convection process by including a fan or possibly other mechanisms that involve liquid or gaseous coolants. The most common active cooling method found in PCs is to couple a fan with a heat-sink, because it is inexpensive and effective. While rotating, a fan helps divert the heat absorbed by the heat-sink from the heat emitting component and dissipate the hot air out of the case. The effectiveness of the cooling process depends on the amount of heat generated by the component and the size and rotational speed of the fan.

In a typical PC, a fan is placed on the CPU and in the power supply. Other common configurations include the placement of fans on the left/back side of the case. Since cold air is denser than hot air, the cold air moves downward while the hot air is pushed upward. Because of this fact, the fan configurations in a PC are typically designed to bring in cold air from the front of the case (at the bottom) while pushing the hot exhaust out from the back (at the top). Fans are either wired to the motherboard through dedicated connections or directly to the power supply with a 3-pin connector. For speed control, fans may be wired to the motherboard or to an external controller with a 4-pin connector. Typical fan speed is approximately a few thousand revolutions per minute (RPM). Fans can be controlled in the OS by software, through OS device manager services, and the computer's BIOS, but the most dominant factor affecting fan speed is the current CPU temperature. Table 2 summarizes the fans found in common PCs.

Table 2. Common PC fans

| Location | Primary role | Optional | RPM |
|---|---|---|---|
| Motherboard (CPU socket) | cools the CPU | No | 800-4000 |
| Front bottom | pulls cold air in from outside | Yes | 800-4000 |
| Rear top | cools power supply and expels exhaust | Yes | 800-4000 |
| Side | cools GPU and drives | Yes | 800-4000 |

## IV. PHYSICAL CHANNEL EXPERIMENT SETUP

In this section, we discuss our experimental setup which is designed to evaluate heat emissions as a viable signaling channel between two PCs. Since the proposed channel is half-duplex, we will evaluate one direction of a communication at a time. Therefore, for the duration of this paper, we refer to the PC that emits the heat as the '*sender*' or '*transmitter*' and the PC that monitors the environmental temperature as the '*receiver*'. A channel configuration reflects the technical and spatial configurations that two communicating PCs may have, directly affecting the channel's performance.

### A. Experiment Methodology

The experiments involved two types of setups: a single PC setup or a setup involving a pair of PCs. With the single PC, we evaluated the thermal properties (e.g., heating and cooling rates) of an active PC under different workloads. In the setup involving two PCs, we evaluated the mutual thermal effects between adjacent computers First, using the single PC setup, we performed a series of CPU workload trials. The objective was to understand the amount of heat that was effectively generated and how quickly this heat dissipates, as well as how these changes registered in the different sensors. For each trial, we generated different workloads and monitored the local heat from the same computer's sensors. Afterwards, using the two PC configurations, we examined the effect of the transmitter's heat on the receiver and its environmental sensors. Several series of trials were performed to analyze the many possible combinations that may affect the channel. For instance, some of the parameters analyzed were the transmitter/receiver's relative distance, chassis type, and relative layout. Lastly, we tested the feasibility of emitting heat from within a virtual machine guest OS. We found that although the physical CPU is controlled by the host OS, the guest OS can use it indirectly as a thermal transmitter.

### B. Experiment Setup

#### 1) Test environment

Tests were conducted in two environments: 1) a regular office room (3 by 4 meters) with standard furniture, no windows, and a closed door, and 2) a larger shared office space (4 by 9 meters) with a small window and an open door. The computers were either placed above or below a table or inside a computer desk tray. An additional experiment was conducted in an open room environment (described later). Environmental temperature during the tests was controlled using the office's air-conditioning system.

#### 2) Hardware

We used two sets of PCs for the experiments. The first set consisted of PCs with a GIGABYTE GA-H87M-D3H motherboard and a quad-core hyper-threaded Intel i7-4790 processor with a built-in Intel HD Graphics 4600 display adapter. Both computers had an identical, regular no-brand tower case, with one fan in the back of the case in addition to the CPU and PSU fans. The chassis had a perforated square area on the left side, in front of the CPU fan, and a perforated region near the back and PSU fans. The second set of PCs consisted of Lenovo m57-6072 computers with an Intel Q35 Express chipset, an Intel core 2-duo processor with an integrated display adapter. The Lenovo 6072 had a small-form-factor case with a perforated front and back panels with a single fan blowing air out of the back. We also conducted tests with large tower case manufactured by Gigabyte from the GZ series with a H77-D3H motherboard, and a quad-core hyper-threaded Intel i7-3770 with a built-in Intel HD Graphics 4000 display adapter. The large case dimensions are 9X48X41 cm. For most of the tests, fan speeds were controlled by the default settings of the motherboard. The impact of environment temperatures on fan speeds was tested using an AK-FC-08-AT Asetek fan controller kit. The virtual machine heat emission experiment was conducted using Oracle VirtualBox. The environmental temperature and heat expansion around the PCs were captured by a Fluke VT04A visual infra-red thermometer and a FLIR T335 thermal imaging camera.

#### 3) Software

In order to generate heat, we performed CPU and GPU stress tests. Our code placed heavy stress on the processors by performing CPU intensive calculations and busy loops. We also used FurMark, a GPU stress tester [22] and prime95, a program for finding Mersenne primes that induces a high CPU load and is popular for CPU benchmarking and stress testing [23]. The computers in the experiments used a background program called HWInfo [24] which sampled and recorded the various thermal sensors and PC fan speeds. The sample rate was 0.5Hz, and all monitored sensors were CPU and motherboard stock sensors. Using the 0.5Hz sample rate, HWInfo in itself utilizes very little CPU time and we validated that it had a negligible effect on the measurements.

#### 4) Layouts and Distances

We conducted tests with different layouts of adjacent pairs of PCs. The layouts reflected common scenarios in which two PCs are placed next to one another in typical office environments. Close proximity between PCs may occur in a shared office shared by two employees with adjacent desks or in a private office in which a single employee has more than one personal computer near his desk. The latter is common in work environments with air-gapped networks since security measures usually require the employee to have a separate computer for each network.

From our observations, in the typical scenario, two desktops are placed side by side (Figure 2) at distances ranging from 1

to 40 cm. Moreover, the PCs were generally positioned with the back panel facing a wall or in the middle of the room. This layout is used in our experiments unless otherwise mentioned, and is referred to as the *parallel layout*. Other layouts tested include: 1) both PCs placed horizontally, one on top of the other, with the transmitting computer placed either on the top or the bottom (i.e., *stacked layout*); 2) two computers placed back-to-back, with the back panels facing each other; this layout reflects the common scenario of two employees sharing an office with desks facing each other (i.e., *face-away layout*), and 3) two computers positioned in some form of angle (i.e., *quadrature layout*). An illustration of the different layouts is given in Appendix A.

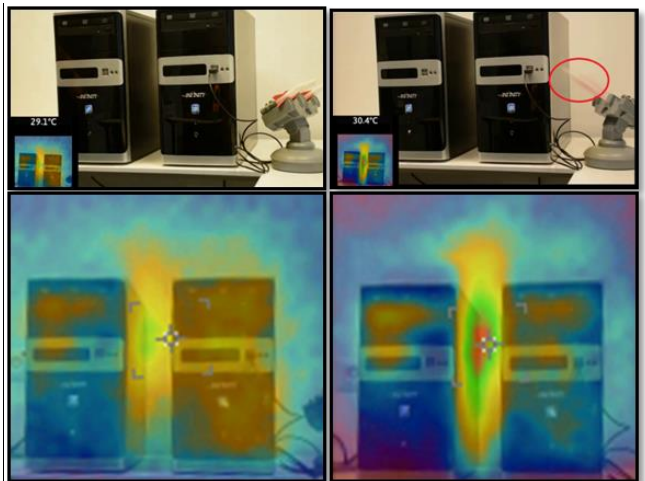

**Figure 2. Two air-gapped PCs positioned in the parallel layout. The left computer transmits a command that instructs the right computer to calibrate & fire a USB game rocket.**

V. PHYSICAL CHANNEL EXPERIMENTAL RESULTS

In this section, we discuss the physical characteristics of the channel obtained from the experiments described in section IV. Based on the experimental results, we developed a basic communication channel (section VI).

A. *Internal Heat Emissions of a PC*

The thermal properties of a PC (i.e., heating-rate versus CPU utilization) are important since they directly affect the signal quality. During our research, we identified three types of sensors that are significantly affected by the environment's temperature: A) the CPU's internal temperature sensor, B) other temperature sensors such as those in the CPU and HD, and C) the fan speed in RPM. As described in previous sections, a PC's workload has a direct effect on the temperature of its components. This subsection examines how and when heat is emitted from within a PC in order to understand how that may affect a thermal communication channel.

1) *Warming and Cooling*

Figure 3 shows a PC's temperature captured by its various thermal sensors over 40 minutes of 100% CPU utilization followed by 40 minutes of nearly 0% utilization. It is apparent that CPU loads have a direct impact on the CPU core's thermal sensors (sensor group A) due to an almost instantaneous increase of about 20°C in the temperature. The motherboard's temperature sensors (sensor group C) showed a +10°C change over 30 minutes. The voltage regulator temperature remained static for the entire duration of the test and was removed from the figures in all further analysis. From a spatiotemporal perspective, the sensors closer to the heat source were more intensely impacted. Note that it takes about 1.5-3 minutes of continuous work for the ambient temperature sensor to increase by 1°C. This rate decreases slightly as temperatures rise but remains constant for a maximum of 10°C above the idle temperature.

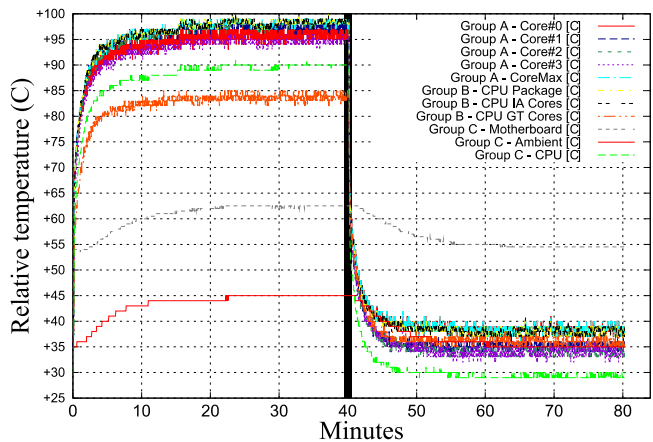

**Figure 3. Temperature readings from various thermal sensors in a desktop PC as influenced by 100% CPU utilization over 40 minutes.**

Once the full CPU utilization stops (minute 40), its internal temperature immediately drops. The ambient case temperature sensor shows that it took about 1-3 minutes for a drop of 1°C, with the rate decreasing slightly as it gets closer to the idle temperature.

In general, we observed that the PC heats up faster near its idle temperature and cools faster near its maximum temperature. This is because at higher temperatures, the fans rotate faster in order to conduct the heat at a faster rate.

2) *Casual Thermal Emissions*

It is important to measure the heat that a PC generates during casual work in order to anticipate its effect on the thermal communication channel. Obviously, a frequently changing PC workload can generate internal heat that interferes with the original heat transfer. Therefore it is clear that the optimal time to emit thermal signals would be when the PC is idle (e.g., during the night). However, in order to demonstrate the negligible affects casual usage has on the channel, we evaluated the system's thermal emissions during typical daily tasks.

Figure 4 shows a PC's internal ambient temperature alongside its CPU utilization and CPU internal temperature during 30 minutes of daily work. As a reference, we used a typical workstation with an Intel i7-4790 processor, running windows 7 64Bit. The user's activities include writing a document using Microsoft Word, browsing several websites concurrently using Google Chrome with multiple tabs, and watching video stream on YouTube. As can be observed from the test, none of the workloads accompanies by a rise in ambient temperature. The thermal level remained essentially the same while the ambient temperature was constant at 32℃. Therefore, it is feasible to use a workstation as a receiver during regular work hours, as long as the local CPU does not have an unusual workload. As can be noticed, the CPU heating sensor is much prone to interference by a user's regular work patterns.

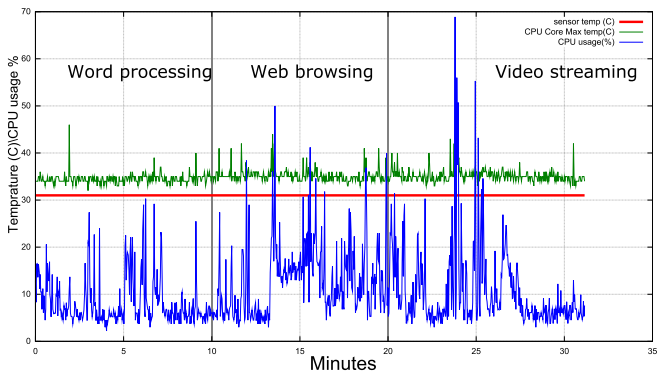

**Figure 4. A PC's internal ambient temperature compared to its CPU internal temperature and utilization during routine use.**

*B. Two Party Thermal Channel*

Thus far, we have shown that a PC's heat level can be controlled by regulating different workloads on CPU. It is known that a PC's internal fans vent the heat exhaust. This heat is emitted to the immediate local vicinity and beyond. In the following subsections we show how the emitted heat affects nearby computer cases, thereby signaling other PCs via their thermal sensors

*1) Sensor Selection*

A Single Input Multiple Output (SIMO) channel correlates a signal received from multiple vantage points, thereby boosting the signal with respect to natural noise and other interferences. It is possible to form a SIMO channel by correlating all of the thermal sensors together. However, for simplicity, we will only view the channel as a single input single output (SISO) channel. Improving the results (e.g., bandwidth) by using a SIMO channel will be addressed in future dedicated research.

A receiver PC on the channel output must monitor its thermal sensors continuously in order to identify changes in the ambient temperature. To maximize the signal quality, we performed experiments in order to identify the sensors which best capture the environmental temperature changes. Based on our experiments, we identified two important aspects of thermal sensors related to receiving signals on the channel's output: 1) the sensor's technical specifications, and 2) its location within the receiver.

Sensor specifications include: 1) sampling rate, 2) accuracy, 3) sample quantization resolution, and 4) value range (minimum and maximum temperature). Most sensors offer a sampling rate higher than required for accurate measurements. A more accurate sensor with higher quantization resolution can better distinguish between subtle changes in temperature level, thereby increasing the channel bit-rate. The range of the sensor must include the minimum and maximum temperatures used by the channel.

The experimental results indicate that the ambient sensor from sensor group B is the best sensor to use in the SISO channel. This is due to its accurate response to environmental temperature changes and the fact that it is relatively unaffected by the sharp CPU spikes caused by casual background processes. Some of the tests results appear in Figure 5.

We note that in cases in which a PC is unequipped with these thermal sensors (e.g., they are taken out or inaccessible), even the thermal sensor of HD may be used to sense the channel's output.

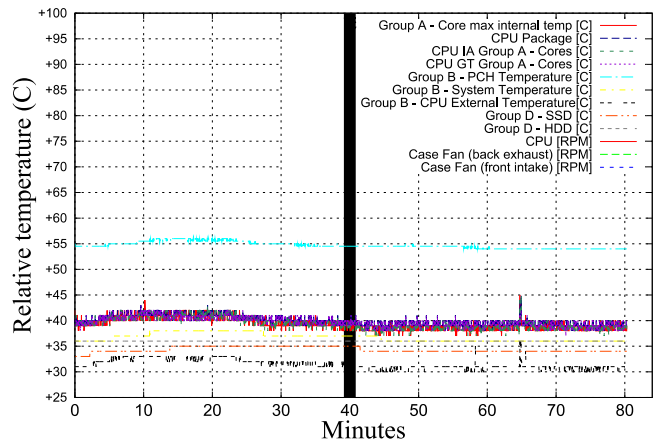

**Figure 5. Temperature readings from various thermal sensors in a desktop PC as externally influenced by a transmitting computer over 40 minutes.**

*2) Distances*

An important part of the channel setup is the distance between the transmitter and receiver. The effect of the heat emitted from a PC's exhaust ports has a limited radius, thereby limiting the maximum channel distance. Beyond that radius, the heat intensity fades and expands across the room. Moreover, in common with all physical communication channels, there is a propagation delay between a channel's inputs (transmission) and outputs (its reception some distance away). The propagation delay is directly correlated to the distance between the two parties. During the experiments we observed that the receiver was not able to sense the change in temperature at a distance of more than 40 cm from the transmitter.

In order to measure and evaluate channel properties, we had the transmitter send a steady signal using consistent heat generation, to a receiver computer. The receiver logged the observed temperatures, each time at an increased distance. In this experiment, we used the common parallel layout and conducted a series of trials at distances of 0, 5, 10, 15, 20, 25, 30, and 35 cm, and the results of the trials are documented in Figure 6. We observed that at short distances, the transmitter could affect the receiver by 1-4℃. For example, when placed next to each other (~0 cm), the heat propagation delay for the first 1℃ increase was three minutes. After 26 minutes the receiver recorded a +4℃ relative change.

At longer distances (e.g., 30-35 cm), we observed that the transmitter could increase the receiver's temperature by at most 1℃. We mention this observation as a constraint for the digital modulation method in section VI.

The experiment clearly demonstrates the relationship between the channel distance and the channel capacity (here represented as the output signal's heat propagation delay). When within a certain radius from the transmitter, there is a near linear relationship: every 1 cm of distance from the receiver increases the heat propagation delay by approximately 0.35 minutes.

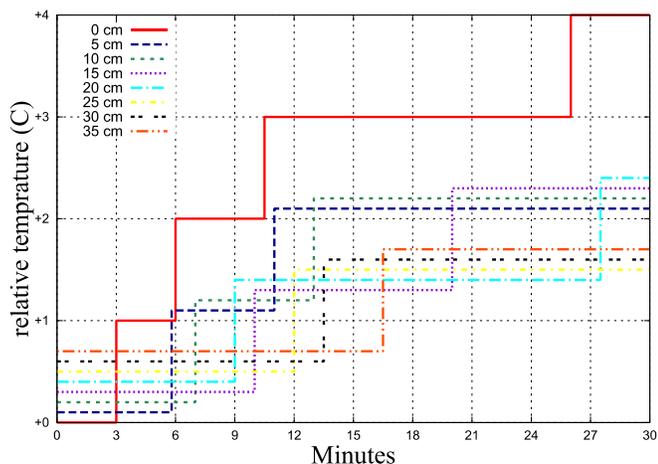
**Figure 6. Relationship between distance and rate of temperature increase in the parallel layout.**

*3) Channel Asymmetry*

When the channel is used as a half-duplex channel, we noticed that the transmission times for each direction were different. Intuitively, this type of asymmetric relation could be caused simply because of the variety of PC cases or different hardware components. However, more interestingly, when using the same type of case and hardware, the channel was still asymmetric. We found, however, that this was based on case orientation: for each common case orientation, the relative spatial locations of the motherboard's heating sources had created different obstacles between them with respect to the other party's sensors. Moreover, we observed that the bottom device benefits from a stacked configuration, since heat rises.

Figure 7 illustrates the asymmetry of the half-duplex channel's capacity between a pair of adjacent PCs in the parallel layout. As can be seen, the heat propagation delay in left-to-right transmissions is smaller than a right-to-left.

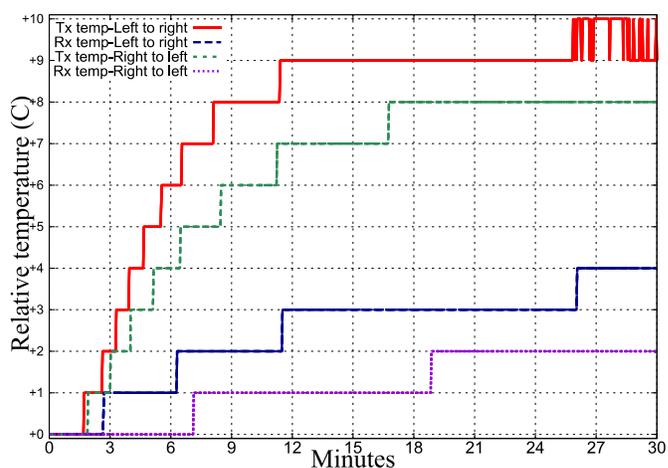
**Figure 7. Channel asymmetrical properties.**

*4) The Effects of Different Orientations*

The attacker is usually unaware of the physical layout of the communicating PCs with respect to one another within the context of the attack scenario (section II), and obviously cannot control the layout. We explore the channel's effectiveness across different possible layouts in order to determine the best and worst case scenarios in term of distances.

Table 3 provides a summary of the results when a transmitter sends a steady signal to the receiver for computers positioned in the stacked layout comparing those with top computer heating to those with bottom computer heading. Our experiment showed that the computers variously have a heating propagation delay of five minutes (transmitter at top) or 12 minutes (transmitter at bottom). As previously explained, the difference is due to the location of the motherboard in relation to the receiver's case: top computer motherboard are at the lower part of the case, closer to the bottom computer.

**Table 3. Properties of the stacked layout.**

|  | With top computer heating | With bottom computer heating |
|---|---|---|
| Heating computer temperature range (°C) | +9 degrees | +10 degrees |
| Adjacent computer temperature range (°C) | +3 degrees | +1 degree |
| Heat propagation delay - first degree increase | 5 minutes | 12 minutes |
| Heating time - to max temperature (heater/adjacent) | 11/20 minutes | 18/12 minutes |
| Pause propagation delay - first degree decrease | 8 minutes | 10 minutes |
| Cooling time - from max to idle temperature. | 20+ minutes | 10 minutes |

When the PCs were positioned in the face-away layout (where the backs of the PCs are facing each other), the heat propagation delay was almost 10 minutes, after which, no further increase was recorded over 40 minutes of heating. However, the cooling time with this layout is less than 3 minutes, which is relatively fast. Notably, no more than a 1℃ increase was observed during 40 minutes of heating. This suggests that the unobstructed intake of cool air from the sides of both cases has a significant cooling effect. Figure 8 details the sensor reading of a test preformed using this layout.

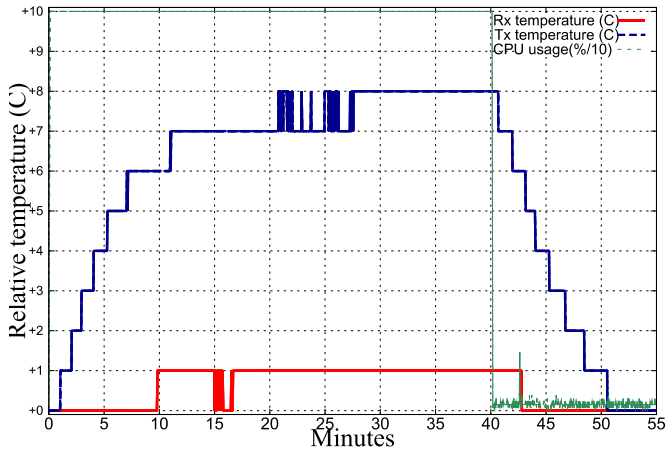

**Figure 8. Heat transfer during 40 minutes of heating, face-away layout.**

Although uncommon, we investigated the scenario where the PCs are in the quadrature layout (where the posterior side of transmitter points to the side of the receiver). From a distance of 6 cm between the transmitter and the receiver, the heat propagation delay is 115 seconds, which is relatively fast. An increase of 4℃ took 11 minutes, and the cooling time was approximately 100 seconds.

*5) Effects of location Relative to Walls or Furniture*

From our observations, the distance of the computers from other objects such as walls and furniture has a significant impact on thermal channels. To better understand their effects, we placed a pair of PCs in the parallel layout on top of a table in the middle of a room. In this scenario, the heat propagation rate for 1℃ was about 25 minutes, shown in Figure 9. This was significantly longer compared with an arrangement in which the computers were 35cm apart and placed near a wall or table. The reason for this difference is that in the middle of the room, the hot air exhaust from the heating computer is completely unobstructed and does not affect the nearby receivers. Nevertheless, locating computers in the middle of the room without surrounding furniture or walls is far less common than being under a table or close to walls. .

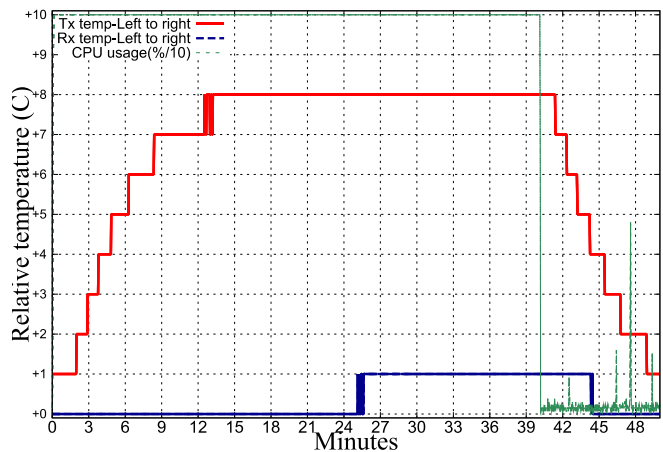

**Figure 9. Heat propagation over 40 minutes, parallel layout in the middle of the room.**

*C. Virtual Machines (VM)*

The virtualization technology has become very popular and accessible in modern IT environments. Workstations in which desktop operating systems host virtual machine guests are commonly used. In this case the host and guests are sharing the physical processor. Thus, overloading the processor from inside the guest may have a limited heat generating effect compared to overloading the physical machine. Figure 10 shows that controlling heat emission is also possible from within virtual machine guests. Tests were conducted using two adjacent computers positioned in the parallel layout. The transmitter side runs as a guest in VirtualBox, with four out of eight logical processors allocated to the virtual machine. Aside from the VM, all other parameters were identical to other tests using physical hosts. The tests indicate that operating from within the VM cause a heat propagation delay of about 3 minutes, only slightly higher than non-VM transmissions. In addition, a maximum temperature increase of 3℃ was observed which is less than the 4℃ increase observed in the experiment using physical hosts. This difference can be attributed to the limitation on the workload imposed by the hypervisor on the processors, primarily due to CPU sharing.

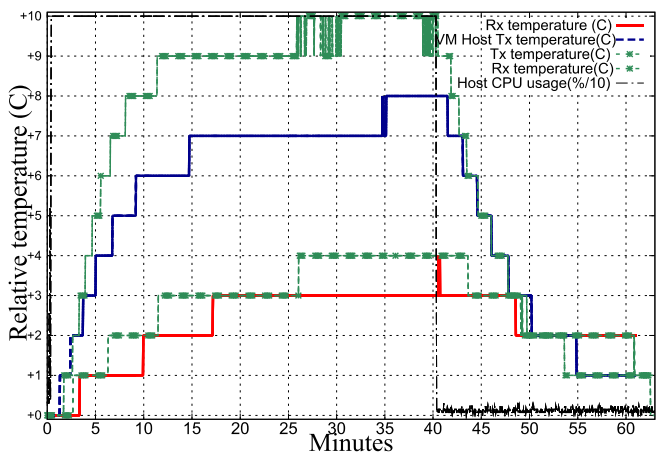

**Figure 10. Propagation delay of virtual machine guest versus physical machine host over 40 minutes.**

VI. COMMUNICATION PROTOCOL

THIS PART IS OMITTED IN THE DRAFT VERSION

THIS PART IS OMITTED IN THE DRAFT VERSION

THIS PART IS OMITTED IN THE DRAFT VERSION     

THIS PART IS OMITTED IN THE DRAFT VERSION    THIS PART IS OMITTED IN THE DRAFT VERSION

## VII. CONCLUSION AND FUTURE WORK

This work introduces a new type of covert channel (codenamed BitWhisper), enabling communication between two air-gapped computers. Our covert channel exploits the thermal radiation emitted by one computer, operating within permissible heat boundaries, to deliver information to a neighboring computer, equipped with standard heat sensors. Our method does not require dedicated or modified hardware, and is based solely on software. We describe a complex yet feasible attack model, based on gradual contamination of computers from isolated networks and public networks, located within one spatial zone, advancing through 'thermal pings' until eventually two spatially adjacent computers belonging to separate networks are found. BitWhisper allows those two air-gapped computers to communicate with one another. We also, implement a working prototype of 'BitWhisper' to evaluate the capabilities and boundaries of the physical layer. A variety of settings, configurations and parameters, with varying distances and several types of target computers were evaluated. We also discuss modulation methods along with the initial handshaking protocol, aimed to deliver narrow-band communication between the two computers, at close distances. Bit-Whisper provides a feasible covert channel, suitable for delivering command and control (C&C) messages, and leaking short chunks of sensitive data such as passwords.

APPENDIX A

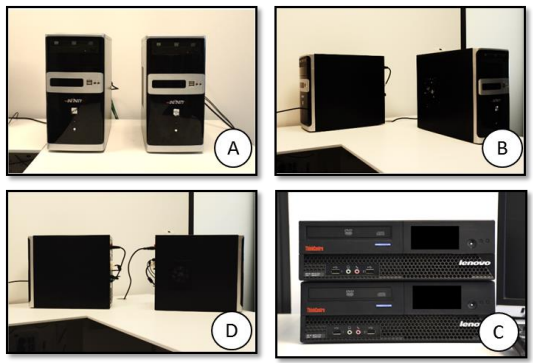

**Layouts illustration: (A) parallel layout; (B) quadrature layout; (C) stacked layout; (D) face-away (back-to-back) layout.**